\documentclass[acmtog,authorversion,nonacm]{acmart}

\usepackage{booktabs}   
\usepackage{subcaption} 
\usepackage{graphicx}   
\usepackage{multirow}

\begin{document}
\sloppy
\title{VVGT: Visual Volume-Grounded Transformer}
\author{Yuxuan Wang}
\authornote{Both authors contributed equally to this research.} 
\affiliation{%
  \department{Department of Mathematics}
  \institution{University of Science and Technology of China}
  \city{Hefei}
  \country{China}
}
\email{wang42@mail.ustc.edu.cn}

\author{Qibiao Li} 
\authornotemark[1] 
\affiliation{%
  \institution{University of Science and Technology of China}
  \city{Hefei}
  \country{China}
}
\email{lqb7@mail.ustc.edu.cn}

\author{Youcheng Cai} 
\authornote{Corresponding author.}
\affiliation{%
  \institution{University of Science and Technology of China}
  \city{Hefei} 
  \country{China}
}
\email{caiyoucheng@ustc.edu.cn}
\author{Ligang Liu}
\affiliation{%
  \institution{University of Science and Technology of China}
  \city{Hefei} 
  \country{China}
}
\begin{abstract}
    Volumetric visualization has long been dominated by Direct Volume Rendering (DVR), which operates on dense voxel grids and suffers from limited scalability as resolution and interactivity demands increase. Recent advances in 3D Gaussian Splatting (3DGS) offer a representation-centric alternative; however, existing volumetric extensions still depend on costly per-scene optimization, limiting scalability and interactivity. We present VVGT (Visual Volume-Grounded Transformer), a feed-forward, representation-first framework that directly maps volumetric data to a 3D Gaussian Splatting representation, advancing a new paradigm for volumetric visualization beyond DVR. Unlike prior feed-forward 3DGS methods designed for surface-centric reconstruction, VVGT explicitly accounts for volumetric rendering, where each pixel aggregates contributions along a ray. VVGT employs a dual-transformer network and introduces Volume Geometry Forcing, an epipolar cross-attention mechanism that integrates multi-view observations into distributed 3D Gaussian primitives without surface assumptions. This design eliminates per-scene optimization while enabling accurate volumetric representations. Extensive experiments show that VVGT achieves high-quality visualization with orders-of-magnitude faster conversion, improved geometric consistency, and strong zero-shot generalization across diverse datasets, enabling truly interactive and scalable volumetric visualization. The code will be publicly released upon acceptance.
\end{abstract}



\begin{teaserfigure}
  \includegraphics[width=\textwidth]{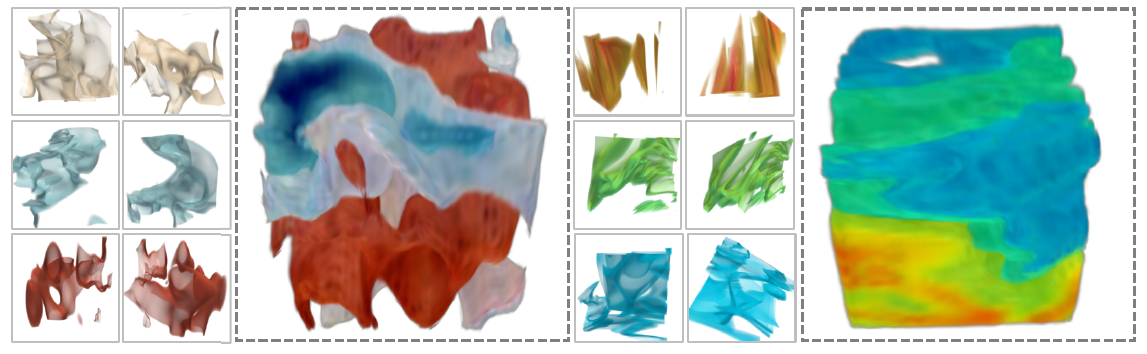}
    \caption{We present VVGT, a general feed-forward framework that takes multi-view images as input and directly converts volumetric data into a 3D Gaussian Splatting (3DGS) representation to enable interactive and explorable volumetric scene visualization. VVGT achieves high-quality visual reconstructions with fine-grained geometric and appearance details. The visualization here demonstrates a composite rendering effect achieved by blending multiple transfer functions (TFs) to highlight distinct volumetric structures. More results can be seen in the accompanying video.}
    \Description{}
  \label{fig:teaser}
\end{teaserfigure}


\maketitle
\section{Introduction}

Volumetric visualization plays a central role in scientific discovery and engineering analysis, supporting applications ranging from medical imaging (e.g., CT and MRI) \cite{klacansky2017openscivis,niedermayr2024novel} to large-scale simulations in fluid dynamics \cite{Jakob20,li2008public}, climate science \cite{athawale2024uncertainty}, and planetary-scale systems~\cite{li2024paramsdrag}. For decades, Direct Volume Rendering (DVR) has been the canonical visualization paradigm. While DVR offers strong generality, its heavy reliance on dense volumetric sampling makes interactive visualization increasingly difficult as data resolution and complexity continue to grow. This fundamental scalability issue persists despite continuous advances in GPU hardware and algorithmic acceleration \cite{niedermayr2024novel}.

Recent progress in explicit radiance representations, most notably 3D Gaussian Splatting (3DGS) \cite{kerbl3Dgaussians}, has revealed a promising alternative: volumetric content can be transformed into a sparse, render-efficient representation that decouples visualization performance from raw data resolution, thereby enabling real-time rendering with high visual fidelity. This shift has sparked growing interest in applying Gaussian-based representations to volumetric visualization, suggesting the possibility of moving beyond DVR toward a representation-centric paradigm. However, existing 3DGS-based volumetric approaches remain fundamentally constrained by their optimization-centric workflow \cite{tang2025ivrgs, dyken2025volumeencodinggaussians}. They typically convert volumetric data into multi-view images and estimate Gaussian parameters via per-scene, iterative optimization. While effective in controlled settings, this process incurs substantial computational cost, scales poorly with data size, and precludes interactive workflows such as rapid exploration, transfer-function editing, or deployment on resource-limited platforms.

In parallel, feed-forward reconstruction approaches \cite{wang2025vggt, jiang2025anysplat, ye2025yonosplat} have shown that explicit 3D representations can be inferred directly via a single network forward pass, eliminating the need for costly optimization. This paradigm shift has reshaped surface-based reconstruction and view synthesis, enabling near-instantaneous geometry inference and strong generalization across scenes. These developments raise a fundamental question for visualization research: 

\textit{Can volumetric visualization move beyond the limitations of DVR and optimization-based pipelines by adopting a feed-forward, representation-first paradigm?}

Answering this question is fundamentally non-trivial. First, a single pixel in volumetric rendering accumulates contributions from multiple spatial locations along a ray, rather than corresponding to a unique surface point. As a result, surface-centric priors—such as per-pixel Gaussian prediction or manifold-constrained geometry—are inadequate by construction. Second, faithfully capturing volumetric structure necessitates reasoning over distributed 3D density and appearance, demanding a tight coupling between multi-view observations and volumetric geometry. These challenges explain why simply accelerating DVR or directly reusing existing feed-forward models is inadequate: the core issue lies in representation and inference, not rendering speed alone.

In this work, we propose Visual Volume-Grounded Transformer (VVGT), a feed-forward framework that establishes a new paradigm for volumetric visualization beyond DVR (see Fig. \ref{fig:teaser}). Rather than rendering directly from voxels or optimizing Gaussians per scene, VVGT learns a direct mapping from volumetric data to a 3D Gaussian Splatting representation, enabling efficient, scalable, and interactive visualization. VVGT consists of two key components. 
First, we introduce a \textbf{Dual-Transformer Network (DTN)} architecture that effectively leverages both 2D appearance information and 3D geometric information. Specifically, we employ a 2D Transformer network to extract appearance features from multi-view images, and a 3D Transformer network to extract geometric features from Gaussians initialized via Variable Basis Mapping (VBM) \cite{li2026variable}. Second, we propose \textbf{Volume Geometry Forcing (VGF)}, a simple yet effective mechanism that encourages multi-view 2D information to be internalized into 3D Gaussian representations. To this end, we introduce an epipolar cross-attention mechanism that aligns the 2D and 3D representations, forcing the initialized Gaussians to learn accurate attributes and thereby enabling high-quality volumetric scene visualization. Extensive experiments demonstrate that VVGT achieves high-quality volumetric visualization with substantially reduced preprocessing cost while delivering competitive or superior visual fidelity compared to DVR and optimization-based 3DGS methods. These results indicate that VVGT constitutes more than an acceleration technique; it represents a representation-level shift for volumetric visualization, enabling interactive exploration, scalable deployment, and practical adoption in real-world scientific and industrial applications.

Our main contributions are summarized as follows:
\begin{itemize}
    \item We present the \textbf{Visual Volume-Grounded Transformer}, the first feed-forward framework that directly maps volumetric data to a 3D Gaussian Splatting (3DGS) representation, enabling efficient Volume-to-Gaussians conversion without time-consuming per-scene optimization.
    
    \item We propose \textbf{Volume Geometry Forcing (VGF)}, a novel epipolar cross-attention mechanism that aligns 2D multi-view appearance features with 3D Gaussian representations, eliminating surface-based assumptions and enabling accurate volumetric 3DGS prediction.
    
    \item We demonstrate that this representation shift enables efficient, generalizable, and interactive volumetric visualization without per-scene optimization, significantly expanding the practical applicability of 3DGS in scientific visualization.
\end{itemize}

\section{Related Work}
\subsection{Volumetric Visualization}
Volumetric visualization is a fundamental topic in computer graphics and scientific computing. Traditional Direct Volume Rendering (DVR) techniques rely on ray casting to integrate optical properties derived from scalar fields via transfer functions~\cite{dvr}. While grid-based representations enable high-fidelity visualization, they incur substantial memory and bandwidth costs when applied to large-scale or time-varying volumetric data. To mitigate these limitations, prior work has explored hierarchical multi-resolution structures, out-of-core streaming, and compression-based rendering techniques~\cite{hiera1,hiera2,brick1,brick2,octree1,oos1,oos2,cp1,cp2}.

Recently, Implicit Neural Representations (INRs) have gained prominence by modeling volumetric density and appearance as continuous functions parameterized by neural networks. Following NeRF~\cite{mildenhall2021nerf}, a series of works such as Plenoctrees~\cite{yu2021plenoctrees}, DVGO~\cite{SunSC22}, TensoRF~\cite{chen2022tensorf}, and Instant-NGP~\cite{muller2022instant} have significantly improved rendering efficiency and memory usage. In scientific visualization, INR-based approaches have been applied to domain-specific tasks, including sparse-view medical imaging~\cite{Cai2024SAXNeRF}, parameter-space exploration~\cite{li2024paramsdrag,Chen2025ExplorableINR}, and compact neural representations using adaptive multi-grid structures~\cite{Wurster2024MultiGridSRN}.

More recently, 3D Gaussian Splatting (3DGS)~\cite{kerbl3Dgaussians} has emerged as an efficient alternative that bridges continuous scene representations and explicit rasterization. Extensions of 3DGS have demonstrated its potential for volumetric visualization, including compressed Gaussian representations for large-scale anatomical data~\cite{niedermayr2024novel}, editable Gaussian volumes for interactive exploration~\cite{tang2025ivrgs}, and semantic, language-driven volume analysis through natural language interfaces~\cite{ai2025nli4volvis}.


\subsection{Feed-forward 3D Reconstruction}
Feed-forward neural networks are widely adopted in 3D reconstruction and novel view synthesis due to their efficiency, scalability, and deterministic inference. Recent spatial foundation models~\cite{liu2021swin,oquab2023dinov2,wu2024point} extend feed-forward paradigms to multi-view settings by jointly reasoning over multiple images and geometric cues. VGGT~\cite{wang2025vggt} processes multiple images simultaneously in a feed-forward manner, while OmniVGGT~\cite{peng2025omnivggt} additionally incorporates auxiliary geometric modalities, such as depth and camera parameters, thereby enabling geometry-aware multi-view inference.

In recent years, several studies have investigated feed-forward Gaussian-based scene modeling. PixelSplat~\cite{charatan2024pixelsplat} infers 3DGS representations from a small number of calibrated images and camera parameters, enabling generalizable novel view synthesis. AnySplat~\cite{jiang2025anysplat} additionally adopts a transformer-based architecture to simultaneously predict Gaussian primitives and their camera poses from uncalibrated image collections. YoNoSplat~\cite{ye2025yonosplat} further improves the feed-forward framework by proposing a mix-forcing training strategy that effectively mitigates training instability and exposure bias. Despite their success, existing feed-forward 3DGS methods are fundamentally designed under surface-centric scene assumptions and therefore cannot be directly applied to volumetric data. In contrast, our work is the first to extend feed-forward 3DGS to volumetric data, overcoming this limitation by jointly modeling multi-view 2D appearance and 3D volumetric geometry.

\section{Preliminary}
\,\,\,\, \textbf{3D Gaussian Splatting (3DGS).} 3DGS~\cite{kerbl3Dgaussians} represents a radiance field using a collection of anisotropic Gaussian primitives in 3D space. Each Gaussian is formally defined as:
\begin{equation}
g(\mathbf{x}) = \exp\left[-\frac{1}{2}(\mathbf{x} - \boldsymbol{\mu})^T {\boldsymbol{\Sigma}}^{-1}(\mathbf{x} - \boldsymbol{\mu})\right],
\end{equation}
where $\boldsymbol{\mu}$ denotes the center of the Gaussian and $\boldsymbol{\Sigma}$ denotes the covariance matrix. The covariance matrix can be factorized as $\boldsymbol{\Sigma} = \mathbf{R}\mathbf{S}\mathbf{S}^{T}\mathbf{R}^{T}$, where $\mathbf{R}$ is a rotation matrix and $\mathbf{S}$ is a scaling matrix. For rendering, the 3D Gaussians are projected onto the image plane using EWA splatting~\cite{zwicker2002ewa}, which enables efficient rasterization-based rendering. After projection, the resulting 2D Gaussians are depth-sorted, and final pixel colors are computed via $\alpha$-blending.

\textbf{Visual Geometry Grounded Transformer (VGGT).} VGGT~\cite{wang2025vggt} is a unified transformer-based framework for 3D reconstruction from multi-view images. Given a batch of input frames, an image encoder~\cite{oquab2023dinov2} extracts corresponding visual tokens $F_i$ for each frame. These tokens are jointly aggregated using global self-attention within a multi-view decoder to produce geometry-aware representations:
\begin{equation}
\{G_i\}_{t=i}^{N} = \operatorname{Decoder}((\operatorname{Global\, SelfAttn}(\{F_i\}_{i=1}^{N})).
\end{equation}

The geometry-aware tokens $G_i$ are passed to prediction heads~\cite{ranftl2021vision} to estimate per-frame camera parameters $c_i$, depth maps $D_i$, and point maps $P_i$:

\begin{figure*}[t]
    \centering
    \includegraphics[width=\textwidth]{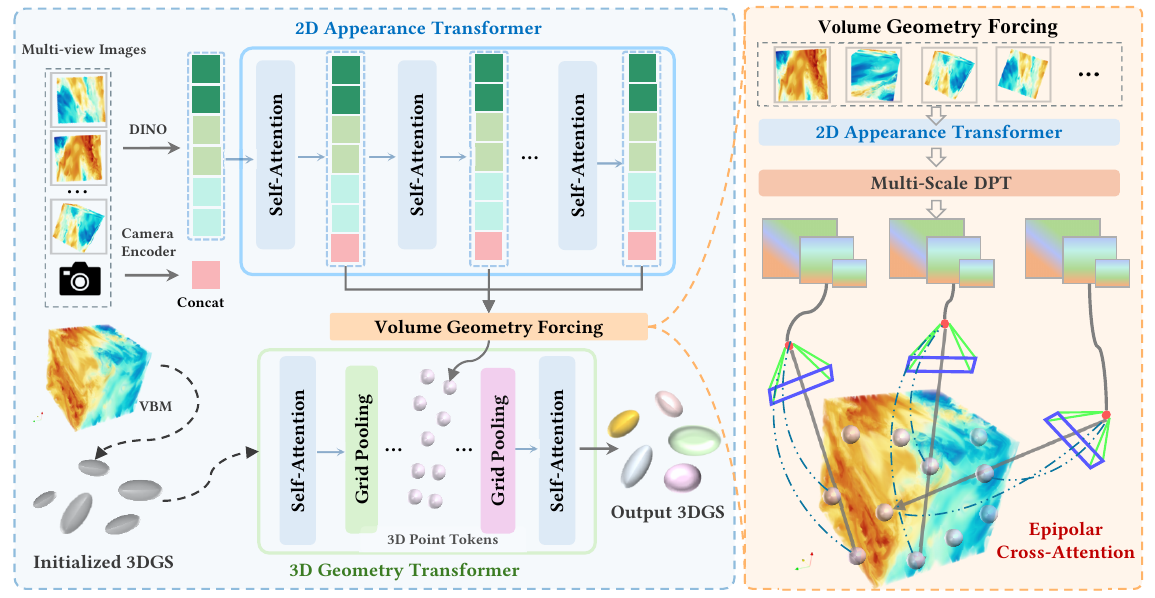} 
    \caption{\textbf{Overview of the VVGT pipeline.} VVGT employs a Dual-Transformer Network with a 2D Appearance Transformer and a 3D Geometry Transformer to jointly model appearance and volumetric geometry. Volume Geometry Forcing (VGF) aligns 2D and 3D features via epipolar cross-attention, enabling accurate Gaussian attribute learning for high-quality volumetric rendering.}
    \label{fig:pipeline}
\end{figure*}

\section{Method}
\label{sec:method}
In this section, we introduce the \textbf{Visual Volume-Grounded Transformer (VVGT)}, a general feed-forward framework for interactive and explorable volumetric scene visualization that distills dense volumetric fields into sparse, high-fidelity 3D Gaussian representations, as illustrated in Fig.~\ref{fig:pipeline}. VVGT consists of two core components. First, we design a \textbf{Dual-Transformer Network (DTN)} that jointly leverages a 2D Transformer to extract multi-view appearance features and a 3D Transformer to capture volumetric geometric features (Sec.~4.2). Second, we introduce a \textbf{Volume Geometry Forcing (VGF)} mechanism that encourages multi-view 2D information to be internalized into 3D Gaussian representations. Specifically, VGF employs an epipolar cross-attention mechanism to align the 2D and 3D representations, forcing the initialized Gaussians to learn accurate attributes and thereby enabling high-quality volumetric scene visualization (Sec.~4.3).
\subsection{Problem Formulation.}
Given a volumetric data $V$ and $N$ calibrated images $\{\boldsymbol{I}\}_{i=1}^N$, VVGT aims to learn a feed-forward network parameterized by $\boldsymbol{\theta}$ that predicts a collection of $K$ anisotropic 3D Gaussian primitives to represent the scene geometry and appearance. Formally, our model learns the mapping, 
\begin{equation}
f_{\boldsymbol{\theta}}: \left(V, \left\{I_i\right\}_{i=1}^N \right) \longmapsto\left\{\left(\boldsymbol{\mu}_k, \sigma_k, \boldsymbol{r}_k, \boldsymbol{s}_k, \boldsymbol{c}_k\right)\right\}_{k=1}^K,
\end{equation}
where $\left(\boldsymbol{\mu}_k, \sigma_k, \boldsymbol{r}_k, \boldsymbol{s}_k, \boldsymbol{c}_k\right)$ denote the parameters of the $k-th$ Gaussian~\cite{kerbl3Dgaussians}, corresponding to its center position, opacity, rotation, scale, and color, respectively. All parameters are inferred from a finite set of input views and can subsequently be used for high-quality visualization from novel viewpoints.

\subsection{Dual-Transformer Network}
Existing feed-forward 3DGS approaches~\cite{charatan2024pixelsplat, jiang2025anysplat, ye2025yonosplat} are built upon a surface-based assumption, in which Gaussians are constrained to surface manifolds and predicted in a pixel-wise manner. In contrast, for volumetric data, a single rendered pixel may correspond to the accumulated contributions of multiple Gaussian primitives along the viewing ray. Therefore, we propose a Dual-Transformer Network (DTN) that jointly processes multi-view 2D images and initialized 3D Gaussians, explicitly modeling both 2D appearance information and 3D geometric information. Specifically, the 2D Appearance Transformer is responsible for extracting view-dependent appearance features from multi-view images, while the 3D Geometry Transformer encodes volumetric geometric priors by aggregating features from initialized 3D Gaussian primitives.

\subsubsection{2D Appearance Transformer} 
\textbf{Architecture.} Following VGGT~\cite{wang2025vggt}, we first partition each input image $I_i$ into non-overlapping patches, which are flattened into a sequence of image tokens. These tokens are concatenated with a learnable auxiliary camera token $e_i$ and fed into a Vision Transformer (ViT) encoder based on the DINOv2 architecture~\cite{oquab2023dinov2}. The resulting features are then processed by a decoder composed of 24 alternating attention layers. Each attention layer consists of two stages. The first stage applies per-frame self-attention, operating independently on tokens from each view to refine local appearance features. The second stage employs global self-attention, where tokens from all views are jointly attended, enabling effective cross-view information exchange and enforcing multi-view appearance consistency.

\textbf{Camera Pose Embedding.} Camera pose information plays a critical role in scene reconstruction and rendering. Unlike generic 3D reconstruction settings where camera poses must be estimated, volumetric visualization typically assumes calibrated cameras, making accurate pose information readily available. We therefore explicitly encode camera pose into the network to guide appearance feature aggregation and facilitate view-aware reasoning.

Specifically, we adopt the camera encoding strategy from OmniVGGT~\cite{peng2025omnivggt}. Given camera parameters $\hat{e}_i$, a dedicated camera encoder transforms them into a feature embedding, which is then injected into the auxiliary camera token:
\begin{equation}
e_i' = e_i + \operatorname{ZeroConv}(\operatorname{Encoder}(\hat{e}_i)),
\end{equation}
where $\operatorname{Encoder}(\cdot)$ denotes the camera pose encoder, and $\operatorname{ZeroConv}(\cdot)$ is a zero-initialized convolution layer that stabilizes training by ensuring a smooth integration of pose information at early stages. The augmented token $e_i'$ is subsequently used throughout the transformer to condition appearance features on explicit camera geometry.

\subsubsection{3D Geometry Transformer.}
\textbf{Initialization.} To overcome the surface-centric assumptions inherent in prior feed-forward 3DGS methods, we explicitly initialize a set of volumetric Gaussian primitives that can jointly contribute along each viewing ray. Inspired by Variable Basis Mapping (VBM)~\cite{li2026variable}, which samples volumetric data in the wavelet domain, we adopt VBM to generate an initial set of Gaussian primitives ${\hat{\mathcal{G}}}_{k=1}^K$, defined as
\begin{equation}
\hat{\mathcal{G}}_k = \left(\hat{\boldsymbol{\mu}}_k, \hat{\sigma}_k, \hat{\boldsymbol{r}}_k, \hat{\boldsymbol{s}}_k, \hat{\boldsymbol{c}}_k\right).
\end{equation}
While VBM provides a structured volumetric initialization, it cannot be directly used for high-quality visualization, as it relies on expensive per-scene optimization to refine Gaussian attributes. We argue that this limitation can be effectively addressed by a feed-forward network that predicts refined Gaussian parameters conditioned on volumetric geometry, enabling efficient and optimization-free volumetric 3DGS generation.

\textbf{Architecture.}
The 3D Geometry Transformer is built upon the PTV3 framework~\cite{wu2024point}, which is specifically designed for processing large-scale point-based representations using attention mechanisms. Each initialized Gaussian $\hat{\mathcal{G}}_k$ is treated as a point primitive and is first mapped to a latent feature through an embedding layer. This is followed by 5 layers of attention blocks interleaved with grid-based downsampling pooling layers~\cite{wu2022point}, which progressively aggregate contextual geometric information. Subsequently, 4 additional layers of attention blocks and upsampling grid pooling layers are applied to restore the feature resolution.

Finally, we adopt a GaussianHead that transforms the PTV3 latent features into refined 3DGS attributes:
\begin{equation}
\left\{\Delta \boldsymbol{\mu}_k, {\sigma}_k, \boldsymbol{r}_k, \boldsymbol{s}_k, \boldsymbol{c}_k \right\}_{k=1}^K =GaussianHead(PTV3(\{\hat{\mathcal{G}}_k\}_{k=1}^K)).
\end{equation}
We predict residual offsets $\Delta \boldsymbol{\mu}_k$ that are added to the initialized Gaussian centers, as the VBM initialization provides reliable spatial locations, while the remaining attributes $(\sigma_k, \boldsymbol{r}_k, \boldsymbol{s}_k, \boldsymbol{c}_k)$ are directly predicted. The resulting Gaussians yield refined volumetric 3DGS representations that are directly suitable for high-quality rendering without requiring per-scene optimization.

\subsection{Volume Geometry Forcing}

To encourage multi-view 2D information to be internalized into 3D Gaussian representations, we propose \textbf{Volume Geometry Forcing (VGF)}, which aligns 2D appearance features with 3D geometric representations. This mechanism forces the initialized Gaussians to learn accurate attributes from multi-view visual evidence, thereby enabling high-quality volumetric scene visualization.

\subsubsection{Multi-Scale Visual Feature Extraction}
To extract 2D visual features suitable for internalization into 3D Gaussian representations, we exploit both multi-layer and multi-scale outputs of the 2D Appearance Transformer. Specifically, we extract token features from $L$ alternating attention blocks of the transformer and feed them into a multi-scale Dense Prediction Transformer (DPT). We augment the DPT with additional decoder heads to obtain multi-scale feature maps $\mathbf{F}_i^{s,l}$, where $s \in \{1,\ldots,S\}$ denotes the scale level and $l \in \{1,\ldots,L\}$ denotes the transformer layer.
In our implementation, we set $S=3$ and $L=4$, forming a feature pyramid that captures both high-level semantic information from deeper layers and fine-grained spatial details from shallower layers. This design explicitly incorporates multi-layer semantic cues and multi-scale spatial context, which are crucial for accurately grounding volumetric geometry in image observations.

\subsubsection{Epipolar Cross-Attention}
We introduce Epipolar Cross-Attention to establish a deterministic and geometry-aware bridge between 2D visual observations and 3D Gaussian representations. This mechanism enables each 3D token to selectively query relevant multi-view image evidence through cross-attention, facilitating effective information transfer from dense 2D appearance to sparse volumetric primitives.
Specifically, for each 3D token $F_k^{pt}$ produced by the 3D Geometry Transformer, we project its 3D position onto all input views using known camera parameters. At the projected locations, we sample the corresponding multi-scale visual features $\{\mathbf{F}_i^{s,l}\}_{i=1}^N$ extracted by the 2D Appearance Transformer, forming paired 3D–2D token sets $(F_k^{pt}, \{\mathbf{F}_i^{s,l}\}_{i=1}^N)$. The epipolar cross-attention operation is formulated as:
\begin{equation}
\tilde{\mathbf{F}}^{pt}_k =
\sum_{i=1}^{N} \sum_{s=1}^{S} \sum_{l=1}^{L}
\operatorname{Softmax}\left(\frac{q_k^{pt} \cdot k_i^{s,l}}{\sqrt{D}} + b_{i,k}^{pt}\right) v_i^{s ,l},
\end{equation}
where $N$ denotes the number of input images, $L$ the number of transformer layers, $S$ the number of feature scales, and $D$ the feature dimension. The 3D token $F_k^{pt}$ generates the query feature $q_k^{pt}$, while the sampled 2D features serve as key–value pairs $(k_i^{s,l}, v_i^{s,l})$. To incorporate geometric consistency across views, we introduce a distance-based attention bias $b_{i,k}^{pt} = -\beta \cdot d_{i,k}^{pt}$ where $d_{i,k}^{pt}$ measures the geometric distance between the 3D token and the $i$-th view. This bias penalizes contributions from geometrically distant views, encouraging the attention mechanism to prioritize visually and spatially relevant observations, similar to geometric priors adopted in prior multi-view learning works~\cite{venkat2023geometry}.


For efficiency considerations, we apply Epipolar Cross-Attention only at the second layer of the 3D Geometry Transformer. We find this to be sufficient, as subsequent transformer layers can effectively propagate and aggregate the fused 2D–3D information while significantly reducing memory and computational overhead.

\subsection{Training Objective}
After applying VGF to inject multi-view 2D appearance information into the 3D Geometry Transformer, the network predicts a refined set of 3D Gaussian primitives through the GaussianHead (Eq.~6). These predicted Gaussians are rendered to produce multi-view images using the 3DGS renderer. Following prior 3DGS-based works, we supervise the rendered images with a combination of pixel-wise and perceptual losses. The overall training objective is defined as:
\begin{equation}
\mathcal{L}_{\text{total}} =\lambda_1 \mathcal{L}_{\ell_1}+ \lambda_2 \mathcal{L}_{\text{SSIM}}+ \lambda_3 \mathcal{L}_{\text{LPIPS}},
\end{equation}
where $\mathcal{L}_{\ell_1}$ measures the pixel-wise absolute difference between the rendered and ground-truth images, $\mathcal{L}_{\text{SSIM}}$ encourages structural similarity, and $\mathcal{L}_{\text{LPIPS}}$ captures perceptual consistency in deep feature space. The weights $\lambda_1$, $\lambda_2$, and $\lambda_3$ balance the contributions of the three terms. This composite objective jointly enforces photometric accuracy, structural coherence, and high-level perceptual fidelity, effectively guiding VVGT to learn accurate volumetric geometry and appearance in a fully feed-forward manner.

\begin{table*}[t]
\centering
\caption{Quantitative comparison on our zero-shot test datasets. We evaluate VVGT against both optimization-based and feed-forward methods. VVGT achieves superior rendering performance over feed-forward methods while remaining competitive with optimization-based methods. In addition, we demonstrate that applying a small number of post-optimization steps yields better rendering quality than optimization-based methods.}
\label{tab:quantitative_comparison}
\setlength{\tabcolsep}{3.5mm}{}{
\begin{tabular}{@{}c|c|cccc|cccc@{}}
\toprule
 \multirow{2}{*}{} & \multirow{2}{*}{Method} & \multicolumn{4}{c|}{24 Views} & \multicolumn{4}{c}{36 Views} \\ \cmidrule(l){3-10} 
 &  & PSNR $\uparrow$ & SSIM $\uparrow$ & LPIPS $\downarrow$ & Time $\downarrow$ & PSNR $\uparrow$ & SSIM $\uparrow$ & LPIPS $\downarrow$ & Time $\downarrow$ \\ \midrule
\multirow{2}{*}{Optimization} & 3DGS & 27.69 & 0.925 & 0.167 & 7min & 28.29 & 0.939 & 0.152 & 7min \\
 & iVRGS & 26.70 &  0.911 & 0.178 & 8min  & 27.28  & 0.929  & 0.162 & 8min \\ \midrule
\multirow{3}{*}{Feed-forward} & NoPoSplat & 10.11 & 0.464 & 0.518 & 2.41s  & 10.26 & 0.442 & 0.522 & 3.62s  \\
 & AnySplat& 13.04 & 0.483 & 0.429 & \textbf{0.94s}  & 13.11 & 0.455 & 0.435 & \textbf{1.60s} \\
 & VVGT& 21.23 & 0.805 &  0.189 & 2.39s  & 22.97  & 0.883 & 0.177 & 4.71s \\ \midrule
\multirow{2}{*}{Post Optimization} & VVGT\_100 & 29.44 & 0.943 & 0.158 & 3.84s  & 29.67 & 0.944 & 0.160 & 6.26s \\
 & VVGT\_200 & \textbf{29.47} & \textbf{0.947} & \textbf{0.150} & 6.69s  & \textbf{29.80}  & \textbf{0.949} & \textbf{0.150} & 8.92s \\ \bottomrule
\end{tabular}}
\end{table*}

\begin{table}[t]
\centering
\caption{Ablation study. We evaluate ablated variants of VVGT, including 3D Geometry Transformer (w/o 3DT), VBM initialization (w/o VBM) and Volume Geometry Forcing (w/o VGF).}
\label{tab:ablation}
\setlength{\tabcolsep}{3.4mm}{}{
\begin{tabular}{lcccc}
\toprule
\textbf{Method} & {PSNR} $\uparrow$ & {SSIM} $\uparrow$ & {LPIPS} $\downarrow$ & Time $\downarrow$ \\
\midrule
w/o 3DT & 12.14 & 0.566 & 0.450 & \textbf{1.60s} \\
w/o VGF & 18.61 & 0.811 & 0.218 & 3.98s \\
w/o VBM & 13.67 & 0.590 & 0.284 & 4.11s \\
\midrule
VVGT & \textbf{22.97} & \textbf{0.883} & \textbf{0.177} & 4.71s \\
\bottomrule
\end{tabular}
}
\end{table}

\section{Experiment}
\subsection{Experiment Setup}
\,\,\,\,\textbf{Datasets.}
To construct our datasets, we use a $4096^3$ temperature field from a rotating stratified turbulence simulation~\cite{Yeung2012} and a $4096^3$ isotropic turbulence simulation~\cite{Lee2015}, both generated via direct numerical simulations. From these volumes, we partition multiple $512^3$ sub-volume scenes. Specifically, we select 100 volume scenes for training and 10 volume scenes for testing. For each volume scene, we use 10 basic transfer functions (TFs) to render images at a resolution of $800\times800$, which are carefully designed to highlight different scalar value ranges and internal structures. All volume datasets are rendered using ParaView with NVIDIA IndeX.

\textbf{Implementation Details.}
We implement our framework in PyTorch. The model is trained using the AdamW optimizer for 40K iterations with a batch size of 8 and an initial learning rate of $1\times10^{-4}$, which is decayed to $1\times10^{-6}$ following a cosine annealing schedule. We train VVGT on 8 NVIDIA RTX Pro 6000 GPUs, each equipped with 96~GB of VRAM, for approximately four days. For 3DGS initialization, we employ VBM~\cite{li2026variable} to construct 150,000 Gaussian primitives for each volume scene.

\textbf{Baselines.}
For comprehensive evaluation, we compare our approach against two categories of methods: (1) optimization-based methods, including \textit{3DGS}~\cite{kerbl3Dgaussians} and \textit{iVRGS}~\cite{tang2025ivrgs}, and (2) feed-forward methods, including \textit{NoPoSplat}~\cite{ye2024NoPoSplat} and \textit{AnySplat}~\cite{jiang2025anysplat}. To evaluate the quality of volumetric visualization, we compute PSNR, SSIM~\cite{wang2004image}, and LPIPS~\cite{zhang2018unreasonable} between the rendered images and the ground truth.


\subsection{Results}
In our experiments, we conduct evaluation on our zero-shot test datasets consisting of 10 volume scenes, each rendered with 10 TFs. During training and inference, we use view settings of 24 and 36 viewpoints, while reserving 10 novel viewpoints for testing. For optimization-based methods, including \textit{3DGS}~\cite{kerbl3Dgaussians} and \textit{iVRGS}~\cite{tang2025ivrgs}, all models are optimized for 30K iterations following their respective standard protocols.

\begin{figure*}[t] 
    \centering
    \includegraphics[height=0.97\textheight, width=\textwidth, keepaspectratio]{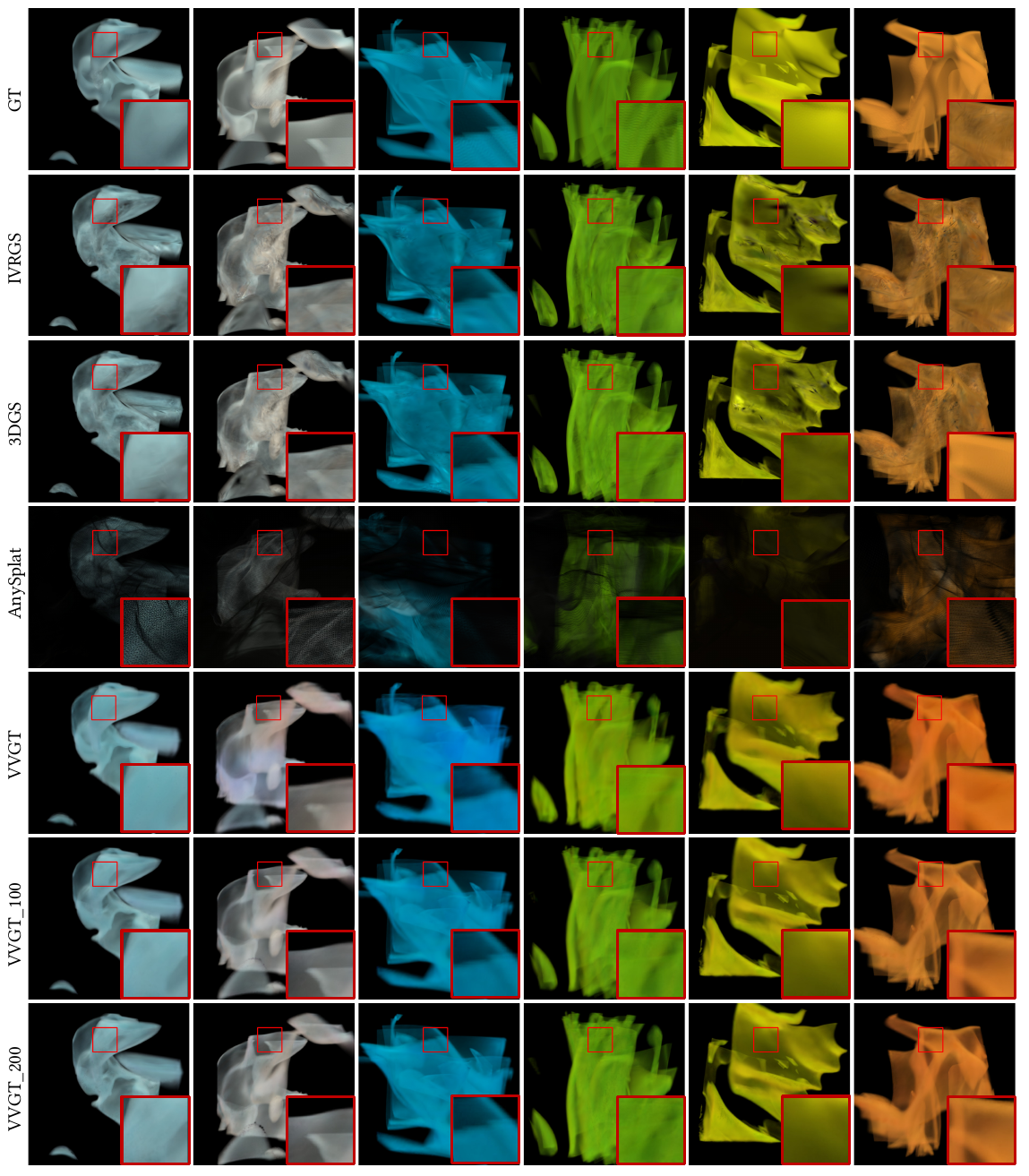}
\caption{Qualitative comparisons using 36 views on our test dataset. Our method demonstrates strong zero-shot performance, outperforming baseline methods in capturing sharp edges and fine-grained details. Moreover, applying a small number of post-optimization steps further improves rendering quality compared to optimization-based methods.}
\label{fig:qualitative_comparison}
\end{figure*}

As shown in Tab.~\ref{tab:quantitative_comparison}, VVGT achieves superior rendering performance on zero-shot datasets compared to recent feed-forward methods, including \textit{NoPoSplat}~\cite{ye2024NoPoSplat} and \textit{AnySplat}~\cite{jiang2025anysplat}, in terms of PSNR, SSIM, and LPIPS. This performance gain can be attributed to two main factors. First, VVGT explicitly incorporates camera pose embeddings, which enable more accurate multi-view reasoning under varying viewpoints. Second, VVGT jointly integrates 2D visual appearance information with 3D volumetric geometric representations, leading to improved consistency and fidelity in rendered results. Notably, VVGT remains competitive with optimization-based methods, despite the latter requiring significantly longer optimization times. Qualitative comparisons in Fig.~\ref{fig:qualitative_comparison} further demonstrate that VVGT produces sharper structures and more coherent volumetric details.

\textbf{Post Optimization.} Although VVGT efficiently performs end-to-end reconstruction of high-quality Gaussian models, further improvements can be achieved through an optional post-optimization step. As shown in Tab.~\ref{tab:quantitative_comparison} and Fig.~\ref{fig:qualitative_comparison}, we demonstrate that applying only 100 steps of post-optimization (taking less than 5 seconds) yields better rendering quality than optimization-based methods even after 30K optimization steps, highlighting the efficiency and robustness of our approach.

\subsection{Ablation Study}

In this section, we analyze the contribution of each individual component in VVGT. All experiments are conducted on our zero-shot test dataset using 36 views. Both quantitative and qualitative results are reported in Tab.~\ref{tab:ablation} and Fig.~\ref{fig:ablation}.

\subsubsection{Effects of 3D Geometry Transformer (w/o 3DT)}
To further evaluate the importance of the proposed 3D Geometry Transformer, we adopt the network architecture of AnySplat~\cite{jiang2025anysplat} and fine-tune it on our constructed volumetric dataset. As shown in Tab.~\ref{tab:ablation} and Fig.~\ref{fig:ablation}, the absence of the 3D Geometry Transformer leads to degraded visual quality in novel-view synthesis. This result demonstrates that per-pixel Gaussian prediction, while effective for surface-centric scenes, is ill-suited to volumetric data visualization.

\subsubsection{Effects of VBM Initialization (w/o VBM)}
VBM initialization provides a reliable geometric prior for volumetric visualization. To assess its impact, we replace the VBM initialization with uniform random sampling of the volume, following the setting in~\cite{tang2025ivrgs}. As shown in Tab.~\ref{tab:ablation} and Fig.~\ref{fig:ablation}, removing VBM initialization results in significantly degraded visualization quality, as the 3D Geometry Transformer struggles to effectively refine geometry without a meaningful volumetric prior.

\subsubsection{Effects of Volume Geometry Forcing (w/o VGF)}
\begin{figure*}[t] 
  \centering
  \includegraphics[width=\textwidth]{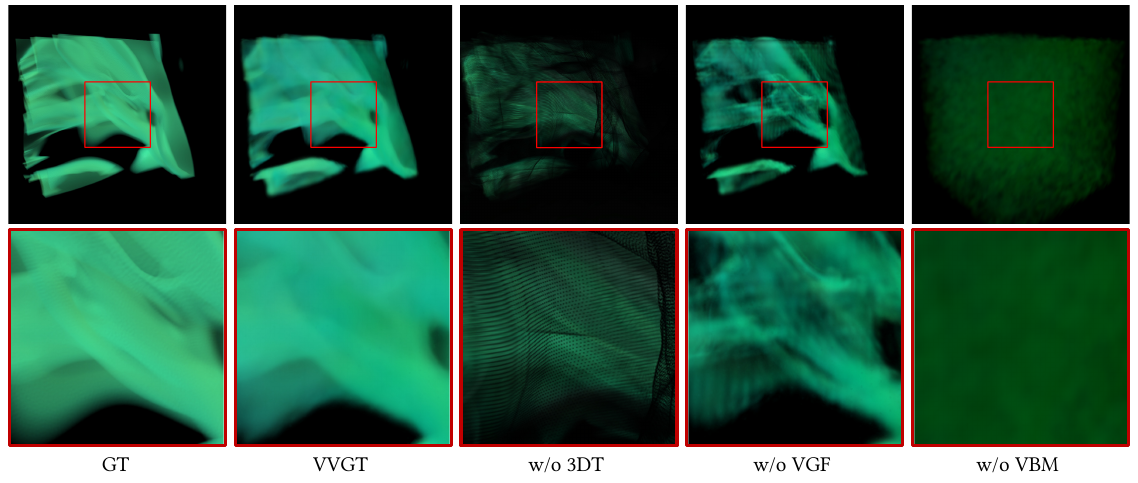}
  \caption{Ablation study of the 3D Geometry Transformer, VBM initialization, and Volume Geometry Forcing. The visual results demonstrate the importance of these modules, as removing any of them leads to a noticeable degradation in visual quality.}
\label{fig:ablation}
\end{figure*}

To demonstrate the effectiveness of Volume Geometry Forcing (VGF), we remove the 2D Appearance Transformer and the Epipolar Cross-Attention module, relying solely on the 3D Geometry Transformer to produce the final results. As shown in Tab.~\ref{tab:ablation} and Fig.~\ref{fig:ablation}, removing VGF prevents the model from learning high-quality 2D appearance information, which in turn leads to blurred structures and inferior visual fidelity in the rendered results.

\section{Conclusion}
We introduced VVGT, a feed-forward framework that introduces a representation-first approach for volumetric visualization that extends beyond traditional DVR methods. By directly mapping volumetric data to a 3D Gaussian Splatting representation, VVGT effectively decouples visualization performance from dense voxel resolution while obviating the need for costly per-scene optimization. This enables near-instantaneous conversion from volumetric data to 3D Gaussian representations while maintaining high visual fidelity and strong geometric consistency. At the core of VVGT are a Dual-Transformer architecture, which jointly reasons over multi-view appearance and volumetric geometry, and Volume Geometry Forcing, which anchors 2D visual observations within distributed 3D Gaussian primitives without dependence on surface-based assumptions. Together, these components address the intrinsic challenges of volumetric rendering—where pixels integrate contributions along rays—and enable feed-forward volumetric 3DGS prediction, achieving feasibility previously unattainable. 

Extensive experiments demonstrate that VVGT achieves high-quality visualization, strong zero-shot generalization, and conversion that is orders of magnitude faster than optimization-based pipelines. These results indicate that VVGT represents not merely an acceleration technique but a conceptual shift in volumetric data representation and visualization. We believe this work establishes a scalable and practical foundation for learning-based volumetric visualization, with broad implications for interactive scientific analysis and real-world deployment.

\textbf{Limitations and Future Work.}
Despite its effectiveness, VVGT has several limitations that highlight promising avenues for future investigation. In this paper, we primarily evaluate VVGT on simulated physical field volumes, which provide controlled settings for validating volumetric structure, rendering fidelity, and representation generality. While these datasets already exhibit diverse volumetric characteristics and transfer-function variations, they do not fully encompass the scale and heterogeneity of large, real-world medical imaging datasets, including CT and MRI scans.

Extending VVGT to high-resolution medical volumes poses additional challenges, such as increased network capacity requirements, higher memory demands, and prolonged training durations to faithfully model fine-grained anatomical details. Addressing these challenges will likely require architectural scaling, more efficient volumetric tokenization strategies, and substantially greater computational resources. Importantly, these challenges are orthogonal to the core representation and inference paradigm proposed in this work and do not compromise the validity of VVGT as a feed-forward volume-to-Gaussians framework.
We view the application of VVGT to large-scale medical imaging as both a natural and impactful subsequent direction. Incorporating real-world CT and MRI datasets, domain-specific priors, and task-driven objectives (e.g., diagnostic or segmentation-aware visualization) represents a significant avenue for future investigation, and the representation-first design of VVGT provides a strong foundation for such extensions.

\bibliographystyle{ACM-Reference-Format}
\bibliography{acmart}

\end{document}